\documentclass[12pt]{article}
\usepackage{graphicx}
\usepackage[cp1251]{inputenc}

\begin{document}
 \noindent {\footnotesize\it Astrophysical Bulletin, 2020, Vol. 75, No 3}
 \newcommand{\dif}{\textrm{d}}

 \noindent
 \begin{tabular}{llllllllllllllllllllllllllllllllllllllllllllll}
 & & & & & & & & & & & & & & & & & & & & & & & & & & & & & & & & & & & & & \\\hline\hline
 \end{tabular}

  \vskip 0.5cm
  \centerline{\bf\large Estimation of the Gould Belt scale height from T\,Tauri}
  \centerline{\bf\large type stars in the Gaia DR2 catalogue}
  \bigskip
  \bigskip
  \centerline
 {V.V. Bobylev\footnote [1]{e-mail: vbobylev@gaoran.ru}
       and
  A.T. Bajkova}
  \bigskip

  \centerline{\small\it Central (Pulkovo) Astronomical Observatory, Russian Academy of Sciences,}

  \centerline{\small\it St.-Petersburg, 196140 Russia}
 \bigskip
 \bigskip
 \bigskip

 {
{\bf Abstract}---We analyze the spatial and kinematic properties of a large sample of young T Tauri type stars
in a 500 pc radius solar neighborhood, closely related to the Gould belt. The following parameters of the
exponential density distribution have been determined: the average $(z_G)_\odot=-25\pm5$~pc and scale height
$h_G=56\pm6$~pc. We propose a method of excluding background stars from the samples, which are located
at large heights with respect to the symmetry plane of the Gould belt. We discovered that the expansion
effect for the entire star system, $K_G=6\pm1$ km s$^{-1}$ kpc$^{-1}$, is determined mainly by the dynamics of the Scorpius–Centaurus association. We show that the angular velocity of the residual intrinsic rotation of
the Gould belt can reac $\Omega_G=6.9\pm0.2$ km s$^{-1}$ kpc$^{-1}$ and that this rotation is opposite to the galactic
rotation.
  }

 \subsection{INTRODUCTION}
The Gould Belt is a giant stellar-gaseous complex closest to the Sun (Efremov 1989; Frogel and
Stothers 1977; Olano 2001; P\"oppel 1997; Torra et al. 2000). A giant neutral hydrogen cloud called
the Lindblad ring (Lindblad 1967) is associated with it, and hosts a large group of nearby OB associations
(de Zeeuw et al. 1999), massive and low-mass stars (Torra et al. 2000), young open star clusters (Bobylev 2006; Piskunov et al. 2006), as well as complexes of dust and molecular clouds (Dame
et al. 1987; Perrot and Grenier 2003; Schlafly et al. 2014).

The second data release of the Gaia space experiment was published in 2018 (Brown et al. 2018; Lindegren
et al. 2018). The Gaia DR2 catalog contains trigonometric parallaxes and proper motions of about
1.3 billion stars. These values were derived based on orbital observations carried out over the course
of 22 months. The average error of determining the trigonometric parallax and both proper motion
components in this catalog depends of the stellar magnitude. For bright stars, $G<15^m,$ parallax errors
lie in an interval of 0.02--0.04 mas, whereas for faint stars, $G=20^m,$ they reach 0.7 mas. Radial velocities
for more than 7 million stars of F--G--K spectral types are determined with an average error of about
1~km s$^{-1}$.

It is impossible to reliably classify stars based only on the Gaia DR2 catalog data. However, it can be
done if these data are supplemented with other photometric and spectral surveys (Marton et al. 2019).
Presently, there are publications dedicated to the selection of young Gaia DR2 catalog stars belonging
to both the OB associations within the Gould belt (Damiani et al. 2019; Ortiz-Le\'on et al. 2018)
and open star clusters (Cantat-Gaudin et al. 2018; Soubiran et al. 2018), and the entire Gould belt as a
whole (Zari et al.2018) (and not only it) (Kounkel and Covey 2019; Marton et al. 2019).

In this work we are interested, in particular, in T Tauri type stars, since the main spatial and kinematic
properties of the Gould belt are determined mainly by analyzing massive stars.

Up until recently, no large samples of T Tauri type stars with reliably measured characteristics existed.
Based on the HIPPARCOS (1997) catalog data, using an example of the closest to the Sun Scorpius--Centaurus OB association, it has been shown that there are no differences in the distribution and kinematics
of high- and low-mass stars of comparable age (Bobylev and Bajkova 2020a; Sartori et al.2003).
It would be interesting to confirm the conclusion of these authors for the entire Gould belt complex. The
necessary data on a large sample of T Tauri type stars selected from GaiaDR2 have recently been published
by Zari et al. (2018). As was shown by a kinematic analysis of these stars, a significant part of the expansion
effect typical of the Gould belt stars can be explained by the influence of the galactic spiral density
wave (Bobylev 2020). The position parameters found from such stars are in a good agreement with the geometric
characteristics of the Gould belt determined earlier (Bobylev 2020).

The aim of this work is to study spatial and kinematic characteristics of the Gould belt based on the
newest data on T Tauri type stars. We use a sample free of background stars to determine the density
distribution parameters in the Gould belt, and also estimate a series of kinematic parameters using the
Oort-Lindblad model.

 \section{THE DATA}
In this work we use the catalog of Zari et al. (2018), which contains T Tauri type stars selected
from the GaiaDR2 catalog by their kinematic and photometric data. These stars are located not further
than 500 pc from the Sun, since a limitation of $\pi<2$~mas was imposed on the sample radius.
The objects were selected based on their proper motions by analyzing a smoothed distribution of
points on the plane $\mu_\alpha\cos\delta\times\mu_\delta$ taking into account the limits of the transverse velocity of the star $4.74r\sqrt{(\mu_\alpha\cos\delta)^2+\mu^2_\delta}<40$~km s$^{-1}$, where the
distance $r$ is computed from the parallax $r=1/\pi$.

The catalog includes three sub-samples of T Tauri type stars as follows:

a) pms1, which includes 43 719 stars, located inside the most outer contour constructed by smoothing
the points on the $\mu_\alpha\cos\delta\times\mu_\delta$  plane, and therefore, this samples contains the most background objects in comparison with the other two;

b) pms2, containing 33 985 stars, located inside the second contour on the $\mu_\alpha\cos\delta\times\mu_\delta$ plane;

c) pms3, with 23 686 stars, located inside the third contour, and therefore they are the most likely
members of the kinematic group (Gould belt).

As was shown by Zari et al. (2018), the stars in all the listed samples, pms1, pms2 and pms3, have a
close spatial connection with the Gould belt. In this work, we consider stars with relative trigonometric
parallax errors less than 15\%.

 \section{METHODS}
 \subsection{Coordinate expressions}
We use Cartesian coordinates centered on the Sun, where the $x$-axis points to the galactic center, the $y$-axis is aligned with the direction of galactic rotation and the $z$-axis is directed towards the north
pole of the Galaxy. We then have $x=r\cos l\cos b,$ $y=r\sin l\cos b$ and $z=r\sin b.$ The heliocentric distance to the star can be computed using the star’s parallax $\pi$, $r=1/\pi$. If $\pi$ is in milliseconds of arc, then the distance is in kpc.

The conversion of equatorial coordinates to galactic was done in accordance with the following expressions:
 \begin{equation}
 \begin{array}{lllll}
 \sin b=\cos\delta\cos\delta_{\scriptscriptstyle GP}\cos(\alpha-\alpha_{\scriptscriptstyle GP})
         +\sin\delta\sin\delta{\scriptscriptstyle GP},\\
 \sin(l-l_\Omega)\cos b=\sin\delta\cos\delta{\scriptscriptstyle GP}
          -\cos\delta\sin \delta_{\scriptscriptstyle GP}\cos(\alpha-\alpha_{\scriptscriptstyle GP}),\\
  \cos(l-l_\Omega)\cos b=\cos\delta\sin(\alpha-\alpha_{\scriptscriptstyle GP}),
 \label{aldl-lb}
 \end{array}
 \end{equation}
where $\alpha_{\scriptscriptstyle GP}=192^\circ.85948,$ $\delta_{\scriptscriptstyle GP}= 27^\circ.12825$ are the coordinates of the north galactic pole (GP) and $l_\Omega=32^\circ.93192$ is the galactic longitude of the ascending node. Numerical values for these parameters for the J2000.0 epoch are recommended by the HIPPARCOS consortium (Perryman et al. 1997). The reverse conversion of galactic coordinates into equatorial is done using the following formulas:
 \begin{equation}
 \begin{array}{lllll}
 \sin \delta=\sin b\sin\delta_{\scriptscriptstyle GP}
         +\cos b\sin(l-l_\Omega)\cos\delta_{\scriptscriptstyle GP},\\
 \cos(\alpha-\alpha_{\scriptscriptstyle GP})\cos\delta=\sin b\cos\delta{\scriptscriptstyle GP}
         -\cos b\sin(l-l_\Omega)\sin\delta_{\scriptscriptstyle GP},\\
 \sin(\alpha-\alpha_{\scriptscriptstyle GP})\cos\delta=\cos b\cos l.
 \label{lb-aldl}
 \end{array} \end{equation}
For a correct determination of the quarter of the sought angle one needs all three formulas from both
the expression groups, (1) and (2).

In this work we want to change over to a coordinate system connected with the Gould belt symmetry
plane, which we will denote by a prime symbol, for example, $l'$ or $z'.$ One may also encounter the need
to change back to the standard, non-prime galactic ordinate system. Expressions (1) and (2) would
allow one to carry out these tasks.

Following Bobylev (2020), for the Gould belt, we use the following coordinates of the GB north pole of
its symmetry plane (large ring):
\begin{equation}
 \begin{array}{lllll}
      l_{GB}=208^\circ.0, \\
      b_{GB}= 78^\circ.0, \\
  l_{\Omega}=298^\circ.0.
 \label{lb-GB}
 \end{array} \end{equation}
Let the initial coordinates of a star be given in the equatorial coordinate system. In this case, in order to
change to the coordinate system connected with the Gould belt symmetry plane by using expressions (1),
we need to know the equatorial Gould belt coordinates, which can be computed using formulas (2):
 \begin{equation}
 \begin{array}{lllll}
  \alpha_{GB}=179^\circ.35316, \\
  \delta_{GB}= 27^\circ.51156, \\
  l'_{\Omega}=269^\circ.35316.
 \label{ALDL-GB}
 \end{array} \end{equation}
If the initial coordinates of the stars are given in the galactic coordinate system, as in the catalog of Zari
et al. (2018), then in order to switch to the coordinates connected with the Gould belt symmetry plane one
should use $l$ and $b$ together with the pole coordinates (3) instead of $\alpha$ and $\delta$ in expressions (1). We will then have $l'$ and $b'$ in the left-hand parts of the equations.

 \subsection{Exponential density distribution}
In the case of an exponential density distribution the histogram of the distribution of stars along the $z$-axis is described by the following expression:
 \begin{equation}
  N(z)=N_0 \exp \biggl(-{|z-z_\odot|\over h} \biggr),
 \label{exp}
 \end{equation}
where $N_0$ is the normalization coefficient, $z_\odot$ --- the average value computed from the $z$-coordinates of the samples stars, which reflects the known fact of the Sun being located above the galactic plane, $h$ --- the vertical scale. In this work we use expression (5) for the distribution of stars in the Gould belt in the
coordinate system related to its symmetry plane, i.e., the primed coordinate system.

 \subsection{Forming the residual velocities}
When forming the residual velocities we first of all take into account the peculiar velocity of the Sun
relative to the local standard of rest: $U_\odot,$ $V_\odot$ and $W_\odot$.  The diameter of the considered region in the vicinity of the Sun is 1 kpc, and we therefore also need to consider the influence of the differential rotation of the Galaxy. The expressions used to account for these two effects have the following form:
 \begin{equation}
 \begin{array}{lll}
 V_r=V^*_r-[-U_\odot\cos b\cos l-V_\odot\cos b\sin l-W_\odot\sin b\\
 +R_0(R-R_0)\sin l\cos b\Omega^\prime_0
 +0.5R_0(R-R_0)^2\sin l\cos b\Omega^{\prime\prime}_0],
 \label{EQU-1}
 \end{array}
 \end{equation}
 \begin{equation}
 \begin{array}{lll}
 V_l=V^*_l-[U_\odot\sin l-V_\odot\cos l-r\Omega_0\cos b\\
 +(R-R_0)(R_0\cos l-r\cos b)\Omega^\prime_0
 +0.5(R-R_0)^2(R_0\cos l-r\cos b)\Omega^{\prime\prime}_0],
 \label{EQU-2}
 \end{array}
 \end{equation}
  \begin{equation}
 \begin{array}{lll}
 V_b=V^*_b-[U_\odot\cos l\sin b+V_\odot\sin l \sin b -W_\odot\cos b\\
 -R_0(R-R_0)\sin l\sin b\Omega^\prime_0
 -0.5R_0(R-R_0)^2\sin l\sin b\Omega^{\prime\prime}_0],
 \label{EQU-3}
 \end{array}
 \end{equation}
where the quantities $V^*_r,V^*_l,V^*_b$ on the right-hand side of the equations are the initial, uncorrected velocities, and $V_r,V_l,V_b$ in the left-hand part are the corrected velocities which we can use to compute
the residual velocities $U,V,W$ based on expressions (10), $R$ is the distance of the star to the rotation axis
of the galaxy:
$$
R^2=r^2\cos^2 b-2R_0 r\cos b\cos l+R^2_0.
$$
We take the distance R0 to be equal to $8.0\pm0.15$ kpc (Camarillo et al. 2018). We adopt the specific values
of the Sun’s peculiar velocity from Sch\"onrich et al. (2010):
$(U_\odot,V_\odot,W_\odot)=(11.1,12.2,7.3)$ km s$^{-1}$. We use the following kinematic parameters:
$\Omega_0=28.71\pm0.22$~km s$^{-1}$ kpc$^{-1}$,
$\Omega^{'}_0=-4.100\pm0.058$~km s$^{-1}$ kpc$^{-2}$ and
$\Omega^{''}_0=0.736\pm0.033$~km s$^{-1}$ kpc$^{-3}$, 
where $\Omega_0$ is the angular velocity of galactic rotation at distance $R_0$, and the $\Omega^{\prime}_0$ and $\Omega^{\prime\prime}_0$ parameters are the corresponding derivatives of this angular velocity.
The values of these parameters were determined by Bobylev and Bajkova (2019) from an analysis of
young open clusters with proper motions, parallaxes, and radial velocities computed using the GaiaDR2
catalog data.

The components $V_r, V_l$ and $V_b$ are used to compute the spatial velocities $U,V,W,$ where velocity $U$ is
directed from the Sun to the center of the Galaxy, $V$ is in the direction of galactic rotation and $W$ points to the north galactic pole:
 \begin{equation}
 \begin{array}{lll}
 U=V_r\cos l\cos b-V_l\sin l-V_b\cos l\sin b,\\
 V=V_r\sin l\cos b+V_l\cos l-V_b\sin l\sin b,\\
 W=V_r\sin b                +V_b\cos b.
 \label{UVW}
 \end{array}
 \end{equation}
Let us estimate the radial velocity errors for our sample in order for them to be comparable with the
transverse velocity errors. In the Gaia DR2 catalog, the average parallax errors for bright stars ($G<15^m$)
lie in the 0.02--0.04 mas interval, and reach 0.7 mas for faint stars (G = 20m). Similarly, the proper motion
errors range from 0.05 mas/yr for bright stars ($G<15^m$) to 1.2 mas/yr for faint ($G=20^m$) stars. If we
take a proper motion error of 0.1 mas/yr, then the transverse velocity error at the sample boundary of
0.5 kpc will amount to $4.74\times0.5\times0.1=0.2$ km s$^{-1}$, and in the limiting case when the proper motion errors are 1 mas/yr, the transverse velocity error at the sample boundary will be $4.74\times0.5\times1=2.4$ km s$^{-1}$. Thus, it is advisable to use the radial velocities of stars with random determination errors less than
2.4 km s$^{-1}$.

 \subsection{Kinematic model}
From our analysis of the residual velocities $V_r,V_l,V_b$ we can determine the average group velocity
$U_G,V_G,W_G,$ as well as four Oort constant $A_G,B_G,C_G,K_G$ ($G$~--- Gould Belt) analogs, which in
our case characterize the effects of intrinsic rotation ($A_G$ and $B_G$) and expansion/contraction ($K_G$ and $C_G$) of a sample of low-mass stars closely related to the Gould belt, based on a simple Oort-Lindblad
kinematic model (Ogorodnikov 1965):
 \begin{equation}
 \begin{array}{lll}
 V_r= U_G\cos b\cos l
     +V_G\cos b\sin l
     +W_G\sin b\\
    +rA_G\cos^2 b\sin 2l +rC_G\cos^2 b \cos 2l +rK_G\cos^2 b,
 \label{MOD-1}
 \end{array}
 \end{equation}
 \begin{equation}
 \begin{array}{lll}
 V_l=-U_G\sin l
     +V_G\cos l\\
 +rA_G\cos b\cos 2l-rC_G\cos b\sin 2l+rB_G\cos b,
 \label{MOD-2}
 \end{array}
 \end{equation}
 \begin{equation}
 \begin{array}{lll}
 V_b=-U_G\cos l\sin b
     -V_G\sin l \sin b
     +W_G\cos b\\
    -rA_G\sin b\cos b\sin 2l
 -rC_G\cos b\sin b\cos 2l-rK_G\cos b\sin b.
 \label{MOD-3}
 \end{array}
 \end{equation}
We find the unknown parameters $U_G,V_G,W_G,$ and $A_G,B_G,C_G,K_G$ as a combined solution of a system
of conditional equations (11)--(13) using the least squares method (LSM). We use a system of weights
in the form of
 $w_r=S_0/\sqrt {S_0^2+\sigma^2_{V_r}},$
 $w_l=S_0/\sqrt {S_0^2+\sigma^2_{V_l}}$ and
 $w_b=S_0/\sqrt {S_0^2+\sigma^2_{V_b}},$
where $S_0$ is the ``cosmic'' dispersion,  $\sigma_{V_r}, \sigma_{V_l}, \sigma_{V_b}$ are the error dispersions of the corresponding observed velocities. The value of $S_0$ is comparable with the least square error $\sigma_0$ (the unit weight error) when solving conditional equations in the form (11)--(13).
In this work $S_0$ ranged from 3 km s$^{-1}$ to 8 km s$^{-1}$. We also used the 3$\sigma$ criterion for discarding residual errors. When analyzing stars with measured radial velocities we also imposed a restriction on the absolute value of the residual velocity $\sqrt{U^2+V^2+W^2}<80$ km s$^{-1}$.

 \begin{figure} [t] {\begin{center}
 \includegraphics[width=164mm]{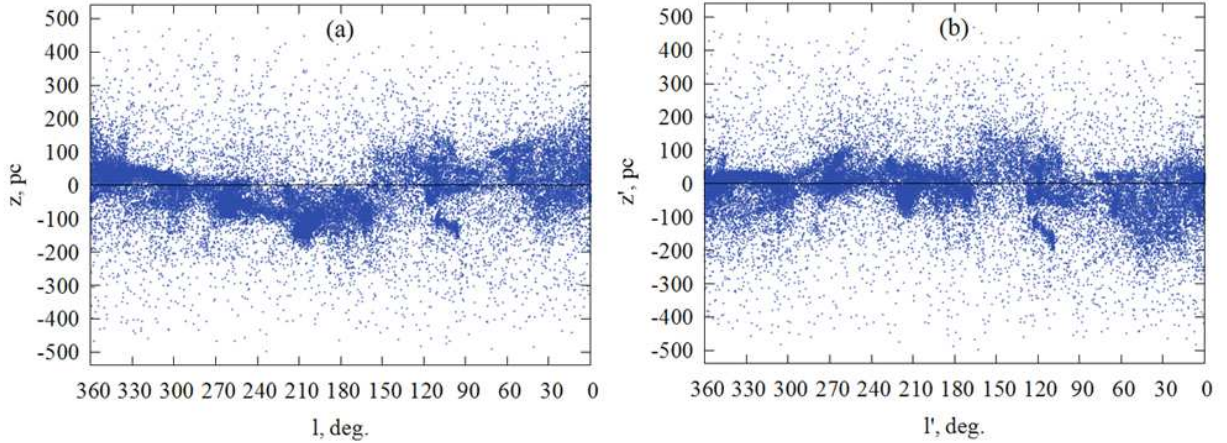}
 \caption{Initial distribution of the pms1 sample stars on a $l-z$ plane (a), and in the new coordinate system $l'-z'$, related to the Gould belt symmetry plane (b).
  }
 \label{f1}
 \end{center} } \end{figure}
 \begin{figure} [t] {\begin{center}
 \includegraphics[width=164mm]{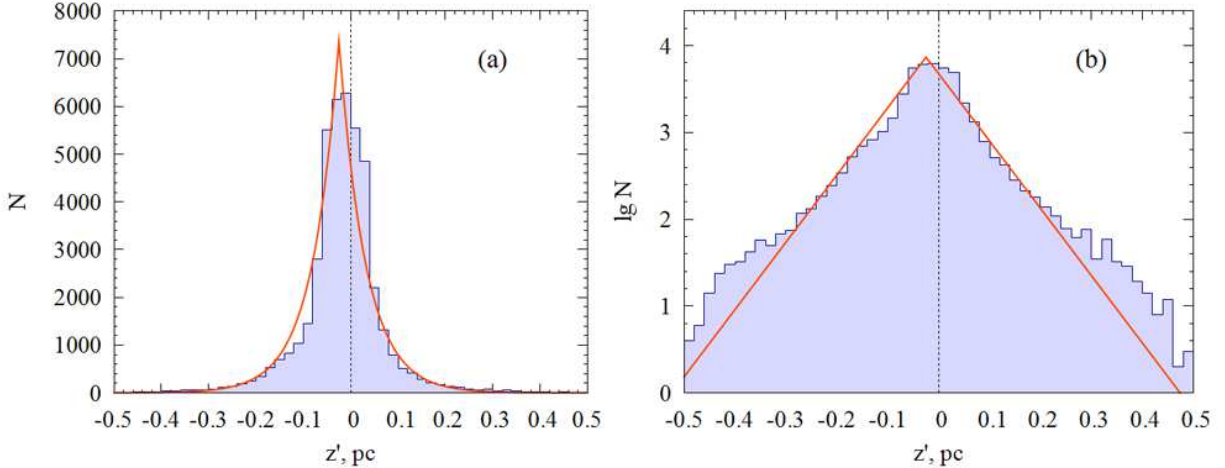}
 \caption{
Histogram of the distribution of pms1 sample stars along the $z$-axis in linear (a) and logarithmic scales (b).
  }
 \label{f-Hist}
 \end{center} } \end{figure}
 \begin{figure} [t] {\begin{center}
 \includegraphics[width=164mm]{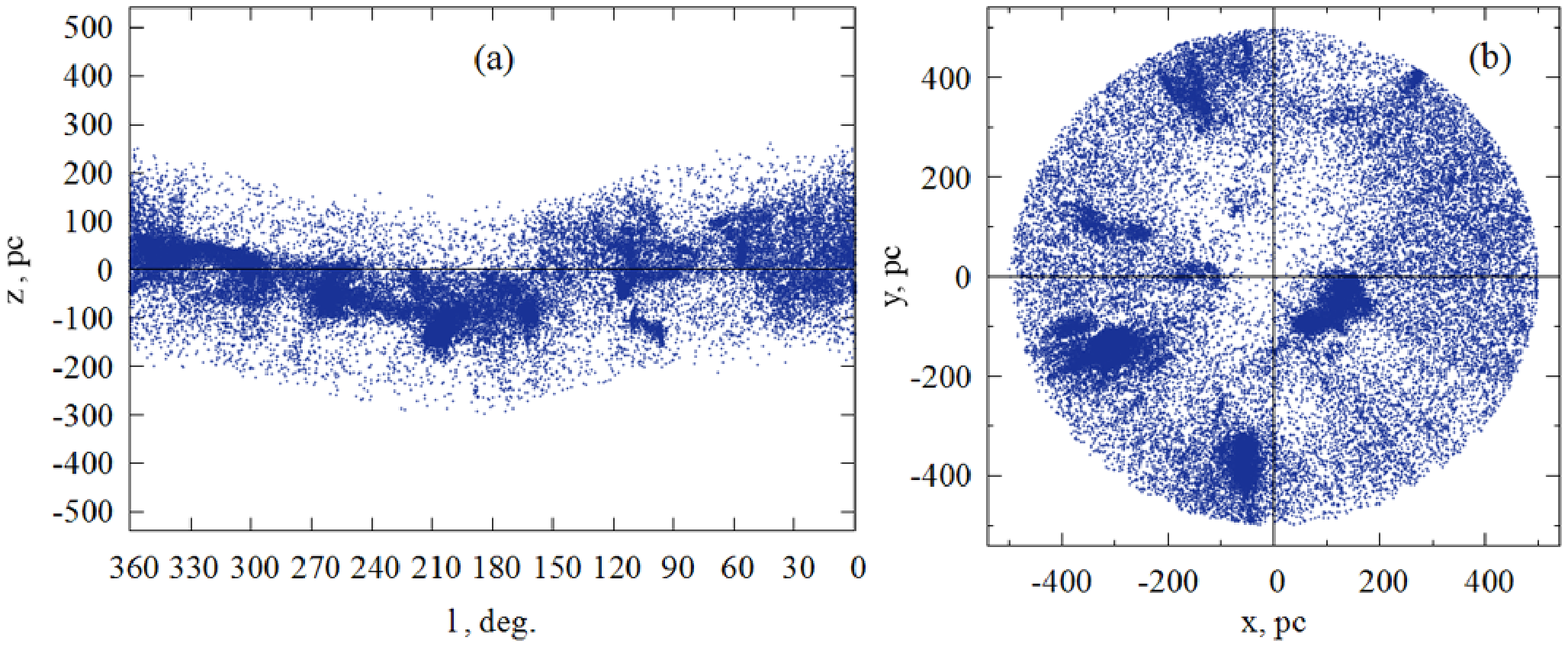}
 \caption{
Distribution of the pms1 sample stars after eliminating the high latitude noise on the $l–z$ plane (a) and projected onto the galactic $xy$ plane (b).
  }
 \label{f-ispr-XY}
 \end{center} } \end{figure}

Using the computed parameters $A$ and $C$, we compute the angle $l_{xy}$ (vertex tilt) according to an
expression proposed by Parenago (1954):
  \begin{equation}
  \tan (2l_{xy+K})={{AK-BC}\over{AB+KC}},
  \label{lxyK}
  \end{equation}
which, in the case of no expansion/contraction (for $K=0$), assumes amore traditional (as in galactic rotation
analysis) form $\tan (2l_{xy})=-C/A.$ In the case of pure rotation, angle $l_{xy}$ points strictly to the kinematic center.

There are several important relations (Ogorodnikov 1965):
  \begin{equation}
   \begin{array}{lll}
    \Omega_0=B-A,\\
        V'_0=B+A,
   \label{Omega}
  \end{array}
 \end{equation}
where, in the case of galactic rotation, $\Omega_0$ is the angular rotation velocity and $V'_0=\partial V_\theta/\partial R$ is the first derivative of the liner rotation velocity $V_\theta$ at point $R=R_0.$

For the angular expansion/contraction velocity $k_0$ and the first derivative of the linear radial expansion/contraction velocity (directed along the radius from the kinematic system center) $V_R$  at point $R=R_0$ we have (Ogorodnikov 1965):
  \begin{equation}
   \begin{array}{lll}
         k_0=K-C,\\
    (V'_R)_0=K+C.
   \label{kkk}
  \end{array}
 \end{equation}
Similar relations are also true for $A_G,B_G,C_G,K_G$ when describing the intrinsic rotation and expansion/contraction of any star system, in particular, that of the Gould belt.

 \section{RESULTS AND DISCUSSION}
 \subsection{Method of eliminating background stars}
Let us first consider one of the three Zari et al. (2018) catalog samples, sample pms1, which contains
the most T Tauri type stars. Fig. 1 shows the initial distribution of pms1 sample stars on a $l-z$ plane, and also in the adopted $l'-z'$ coordinate system, related to the Gould belt plane. A wave is clearly
visible that reflects the fact that the Gould belt is tilted relative to the galactic plane. It is also evident that the sample contains stars with large heights up to 500 pc.

Fig. 2 shows the histogram of the distribution of pms1 sample stars in linear and logarithmic scales.
As is evident from the figure, despite the considerably patchy distribution of stars, on the whole, we
see a satisfactory agreement with the exponential density distribution law. The patchiness, naturally,
is caused by a high concentration of stars in the main associations, which are closely connected with
the Gould belt --- the associations in Lacerta ($l\sim100^\circ$), Cepheus ($l\sim120^\circ$), Cassiopeia ($l\sim130^\circ$), Perseus ($l\sim160^\circ$), Taurus ($l\sim180^\circ$), Orion ($l\sim190^\circ$), Vela ($l\sim270^\circ$) and Scorpius–Centaurus ($l\sim330^\circ$). Fig. 2b clearly shows the wide wings determined by noise. The scale height $h$ for these wings exceeds by a factor of about two the value typical for the vast majority of the remaining stars.

The histogram was used to find the following exponential distribution parameters (5):
 \begin{equation}
 \begin{array}{lllll}
  (z_G)_\odot=-25\pm5~\hbox{pc}, \\
      h_G= 56\pm6~\hbox{pc}, \\
      N_0= 7389.
 \label{exp-data}
 \end{array} \end{equation}
For comparison, we show in Table 1 the results of determining the disk scale $h$ based on the exponential
distribution (5) by different authors using various data. The table contains values of $z_\odot$ related
to field stars. Some authors report $(z_G)_\odot,$ related to, as in our case, the Gould belt. However,
these values are usually not well-determined. For example, $(z_G)_\odot=0\pm2$~pc (Stothers and Frogel
1974), $(z_G)_\odot=-15\pm12$~pc (Elias et al. 2006) or $(z_G)_\odot=7\pm10$~pc (Gontcharov 2019). Comparing
the obtained estimate of $h$ (17) with the data in Table 1 we can conclude that the distribution of low-mass
T Tauri type stars in the Gould belt is very similar to the distribution of massive OB stars.

Let us note the result of Gontcharov (2019), who found a rather large scale height of $h=170\pm40$~pc
for the dust in the Gould belt. One can say that the Gould belt is associated not only with the molecular
and dust clouds (Dame et al. 2001), but also with high-latitude clouds (Schlafly et al. 2014). Bobylev
(2016) has shown that a system of close high-latitude molecular clouds can be approximated by an ellipsoid
with dimensions $350\times235\times140$~pc, the size of the third axis of which differs significantly from
the one usually adopted for a Gould belt ellipsoid $350\times250\times50$~pc. Thus, the dust and gas in the
Gould belt are distributed along the $z$-coordinate significantly higher than stars.

Based on the values of $z_\odot$ and $h$ in the primed coordinate system we can introduce a limit of $|z'-z_\odot|<3h$, thus eliminating the high-altitude noise. We took $|z'-z_\odot|<200$~pc, with enough to spare. Note that with such a limit, a strip will remain in Fig. 1b, parallel to the equator. The results are shown in Fig. 3, where we give the distribution of pms1 sample stars on the $l-z$ plane after noise elimination, as well as the distribution of these stars projected onto the galactic $xy$ plane.

Such a method is used, it seems, for the first time. Usually, when studying Gould belt stars the authors
separate these stars into two layers: equatorial and tilted --- basically, the Gould belt (Elias et al. 2006;
Gontcharov 2019; Stothers and Frogel 1974). When
using our method in the primed coordinate system,
all objects in the tilted layer are eliminated and a
mixed stellar composition remains only in the nodes.
However, due to the fact that our sample contains
a huge number of stars concentrated in associations
clearly belonging to the Gould belt, the contribution
of the remaining background stars should not affect
significantly our determination of the kinematic parameters.

As is evident from Fig. 3b, the distribution of stars has a region of reduced density with the center in
the second galactic quadrant, $l\sim120^\circ$, at distance $r\sim150$~pc. This doughnut-shaped form is typical for the Gould belt (de Zeeuw et al.1999; Perrot and Grenier 2003). The giant neutral hydrogen cloud called
the Lindblad ring has the same shape (Lindblad 1967, 2000). One can also notice a slight condensation of
stars located close to the center of this ``doughnut'' --- the $\alpha$~Per open cluster, whose age amounts to about 35 million years.

 \begin{table}[t]                                     
 \caption[]{\small
Vertical disk scale $h,$ obtained by different authors from young objects in the assumption of an exponential
density distribution
  }
  \begin{center}  \label{t:01}
  \begin{tabular}{|r|c|c|l|l|c|}\hline
  Reference & $z_\odot,$~pc & $h,$~pc  & Sample     \\\hline

  Bobylev and Bajkova (2016b) & $~-5.7\pm0.5$ & $27\pm1$ & 639 masers, ~$R\leq R_0$ \\
  Bobylev and Bajkova (2016b) & $~-7.6\pm0.4$ & $33\pm1$ & 878 HII regions, ~$R\leq R_0$ \\
  Bobylev and Bajkova (2016b) & $-10.1\pm0.5$ & $34\pm1$ & 538 GMC,     ~$R\leq R_0$ \\
  Stothers and Frogel (1974)  & $-24\pm3$ & $ 46\pm7$ & B0--B5, $r<200$~pc    \\
  Reed (2000)                 &  $0$      & $ 45    $ & OB stars    \\
  Bonatto et al. (2006)  & $-15\pm2$ & $ 48\pm3$ & OSC, $<200$~Myr  \\
  Elias et al. (2006)    &$-12\pm12$ & $ 34\pm2$ & OB stars            \\
  Piskunov et al. (2006) & $-22\pm4$ & $ 56\pm3$ & 254 OSC              \\
  Joshi (2007)           & $-17\pm3$ & $ 57\pm4$ & 537 OSC              \\
  Bobylev and Bajkova (2016a) & $-16\pm2$ & $ 45\pm3$ &   OB associations     \\
  Bobylev and Bajkova (2016a) & $-15\pm3$ & $ 49\pm3$ &  187 HII regions, $r<4.5$~kpc \\
  Bobylev and Bajkova (2016a) & $-10\pm4$ & $ 51\pm4$ &  148 Wolf-Rayet stars \\
  Bobylev and Bajkova (2016a) & $-19\pm4$ & $ 61\pm4$ &  90 masers, $r<4$~kpc \\
  Bobylev and Bajkova (2016a) & $-23\pm2$ & $ 70\pm2$ &  246 Cepheids, ${\overline t}\approx~$75~Myr \\
 Gontcharov (2019)            & $-10\pm5$ & $170\pm40$ & Gould belt dust \\\hline
 \end{tabular}\end{center}
 {\small
  }
 \end{table}

 \subsection{Stellar kinematics}
To estimate the effects of intrinsic rotation and expansion/contraction of the Gould belt, we solve a
system of conditional equations (11)--(13) using the least squares method. We search for a solution for
the pms3 sample (which contains the most probable
Gould belt members), free of high-altitude noise. The
results are presented in Table 2, which consists of
two parts: upper and lower. These parts differ by the
method used to solve the kinematic equations. Thus,
for stars with measured radial velocities, each star
gives all three equations (11)--(13). The solutions derived only from such stars are given in the upper part
of the table. The lower part gives solutions obtained in the following way: a star with proper motion gives
two equations: (12) and (13), and if a star has radial velocity it gives all three equations.

In the first column of Table 2 we present the kinematic model parameters. The second column (Step I)
shows the solution obtained from the velocities of stars that were not cleared of any effects. In the
third column (Step II), the Sun’s peculiar velocity relative to the local standard of rest and the differential
rotation of the Galaxy were eliminated from the stellar
velocities. In the fourth column (Step III) we removed
the Scorpius-Centaurus association stars from the
sample, and in the fifth column (Step IV) the solution
was obtained for the sample in the previous step, but with new longitudes of the form $l_{new}=l-l_{xy}$. When removing the Scorpius-Centaurus association stars from the sample, a square with dimensions $x:0-200$~pc and $y:-200-0$~pc is freed.

 \begin{figure} [p] {\begin{center}
 \includegraphics[width=145mm]{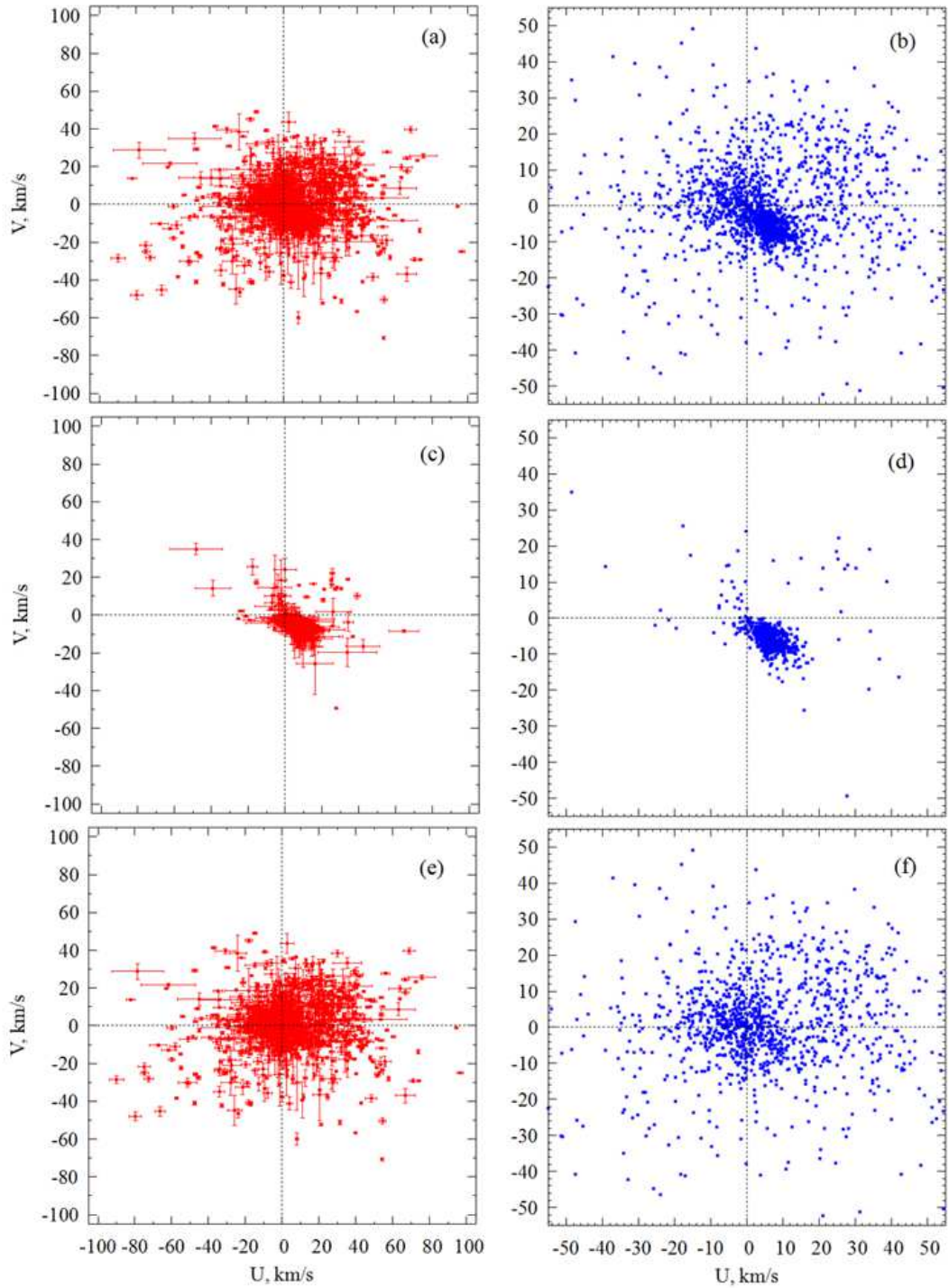}
 \caption{
Residual velocities for 1845 pms3 sample stars (a), the same velocities in another scale (b), residual velocities for 682 Scorpius-Centaurus association stars (c), the same velocities in another scale (d), residual velocities for 1163 pms3 sample stars after the elimination of stars belonging to the Scorpius-Centaurus association (e), the same velocities in another scale (f).
  }
 \label{fig-UV}
 \end{center} } \end{figure}

The decision to remove the Scorpius-Centaurus association stars from the sample is based on the paper
of Bobylev and Bajkova (2020b), where the presence of a very strong association expansion coefficient
was confirmed, $K=39\pm2$~km s$^{-1}$ kpc$^{-1}$. Fig. 4 shows the $UV$-velocities of the pms3 sample stars
before and after removing the Scorpius-Centaurus association stars. As is clearly seen, the distribution
of $UV$-velocities for the Scorpius-Centaurus stars is very compact, it has a specific form related to the
presence of expansion. Such stars are quite numerous, and therefore, their removal has a significant
effect on the nature of the velocity distribution and the estimated Gould belt parameters. Note that here the
$UV$-velocity errors are small. At the same time, in the considered solar vicinity of 500 pc radius, the errors
of transverse velocities $V_l$ and $V_b$  are smaller than those of the radial velocities, since here the parallax and proper motion errors for the stars, taken from the Gaia DR2 catalog, are small. No limits on the velocity were imposed when plotting Fig. 4. The left-hand graphs in the figure show that the absolute values of
the residual velocities rarely exceed 80 km s$^{-1}$.

 \begin{table}[t]
 \caption[]{\small
Parameters of the Oort-Lindblad kinematic model, determined only from the pms3 sample stars with radial
velocities (upper part) and from the full data set (lower part)
 }
  \begin{center}  \label{t:02}   \small
  \begin{tabular}{|c|r|r|r|r|r|r|r|}\hline
  Parameters          &         Step I  &        Step II &       Step III &         Step IV \\\hline
  $N_\star$           &           1845 &           1845 &           1163 &           1163 \\
  $\sigma_0,$ km s$^{-1}$    &            9.2 &            9.1 &           11.5 &           11.5 \\
  $U_\odot/U_G,$ km s$^{-1}$ & $ 5.36\pm0.24$ & $ 5.57\pm0.23$ & $ 5.71\pm0.37$ &  \\
  $V_\odot/V_G,$ km s$^{-1}$ & $11.60\pm0.23$ & $ 0.46\pm0.22$ & $ 2.72\pm0.35$ &  \\
  $W_\odot/W_G,$ km s$^{-1}$ & $ 5.59\pm0.21$ & $ 1.21\pm0.20$ & $ 1.34\pm0.33$ &  \\
  $V_0/(V_0)_G,$ km s$^{-1}$ & $13.95\pm0.22$ & $ 5.72\pm0.23$ & $ 6.46\pm0.37$ &  \\
   $l,$ deg.         & $ 65\pm1$      & $    5\pm2$    & $   25\pm3$    &  \\
   $b,$ deg.         & $ 24\pm1$      & $   12\pm3$    & $   12\pm3$    &  \\
   $A/A_G,$ km s$^{-1}$ kpc$^{-1}$  & $13.55\pm0.92$ & $-2.59\pm0.90$ & $ -0.5\pm1.2$  & $-2.1\pm1.2$ \\
   $B/B_G,$ km s$^{-1}$ kpc$^{-1}$  &$-17.05\pm0.81$ & $-3.52\pm0.80$ & $ -2.3\pm1.1$  & $-2.3\pm1.1$ \\
   $C/C_G,$ km s$^{-1}$ kpc$^{-1}$  & $-4.49\pm0.93$ & $-4.63\pm0.90$ & $ -2.0\pm1.2$  & $-0.0\pm1.2$ \\
   $K/K_G,$ km s$^{-1}$ kpc$^{-1}$  & $ 6.57\pm1.09$ & $ 5.97\pm1.07$ & $  2.0\pm1.4$  & $ 2.0\pm1.4$ \\
 $l_{xy},$ deg.      & $    9\pm2$    &      $-30\pm5$ & $  -37\pm7$    &  $  0\pm7$   \\
 $l_{xy+K},$ deg.    & $         $    &      $ 30\pm8$ & $   32\pm9$    &  $-21\pm9$   \\
  \hline
  $N_\star$           &          23214 &          23214 &          17687 &          17687 \\
  $\sigma_0,$ km s$^{-1}$    &            3.9 &            3.6 &            4.3 &            4.3 \\
  $U_\odot/U_G,$ km s$^{-1}$ & $10.74\pm0.04$ & $ 0.16\pm0.04$ & $ 0.44\pm0.06$ &  \\
  $V_\odot/V_G,$ km s$^{-1}$ & $12.79\pm0.04$ & $-0.75\pm0.03$ & $ 1.26\pm0.05$ &  \\
  $W_\odot/W_G,$ km s$^{-1}$ & $ 6.10\pm0.03$ & $ 0.61\pm0.03$ & $ 1.19\pm0.04$ &  \\
  $V_0/(V_0)_G,$ km s$^{-1}$ & $17.78\pm0.04$ & $ 0.98\pm0.03$ & $ 1.78\pm0.05$ &  \\
    $l,$ deg.         & $  50.0\pm0.1$ & $  282\pm3$    & $   71\pm3$    &  \\
    $b,$ deg.         & $  20.1\pm0.1$ & $   38\pm2$    & $   41\pm2$    &  \\
   $A/A_G,$ km s$^{-1}$ kpc$^{-1}$  & $ 6.29\pm0.11$ & $-8.33\pm0.10$ & $-5.40\pm0.13$ & $-7.16\pm0.13$ \\
   $B/B_G,$ km s$^{-1}$ kpc$^{-1}$  &$-16.49\pm0.09$ & $-2.85\pm0.09$ & $-0.26\pm0.11$ & $ 0.26\pm0.11$ \\
   $C/C_G,$ km s$^{-1}$ kpc$^{-1}$  & $ 4.54\pm0.14$ & $ 4.60\pm0.13$ & $ 4.70\pm0.16$ & $-0.12\pm0.16$ \\
   $K/K_G,$ km s$^{-1}$ kpc$^{-1}$  & $ 6.71\pm0.33$ & $ 5.36\pm0.30$ & $-0.07\pm0.39$ & $-0.07\pm0.39$ \\
  $l_{xy},$ deg.      & $   -18\pm1$   &      $ 15\pm1$ & $   21\pm1$    &   $  0\pm1 $   \\
  $l_{xy+K},$ deg.    & $         $    &      $-17\pm2$ & $   28\pm3$    &   $  7\pm3 $   \\
  \hline
  \end{tabular}\end{center} \end{table}

Let us consider the parameters found in Step I. Since we did not introduce any corrections, the parameters
$A$ and $B$ here describe the rotation of the Galaxy and are influenced by the Gould belt. According
to the data in the upper part of Table 2, the angular velocity of galactic rotation is
$\Omega_0=B-A=-30.6\pm1.2$~km s$^{-1}$ kpc$^{-1}$, and according to the data in the lower part,
 $\Omega_0=B-A=-22.8\pm0.1$~km s$^{-1}$ kpc$^{-1}$.
For example, masers with measured trigonometric parallaxes from the paper of Rastorguev et al. (2017)
were used to derive $\Omega_0=-28.64\pm0.53$~km s$^{-1}$ kpc$^{-1}$ (for model C2, a purely circular disk rotation).

It is notable that the velocities relative to the local standard of rest, determined in Step II, differ significantly. Thus, based on the data in the upper part of Table 2, the stars move with a velocity of
$5.72\pm0.23$ km s$^{-1}$ in the direction $l=5\pm2^\circ$ and $b=12\pm3^\circ$. Here the vector of this velocity lies practically in the Gould belt plane. And according to the data in the lower part of the table, this velocity is very small, and the direction of its vector has a random nature.

Let us consider the solutions given in the last column of Table 2. As is evident from the upper part of the table, the angular residual rotation velocity $\Omega_G=B-A=-0.2\pm1.2$~km s$^{-1}$ kpc$^{-1}$ and the angular expansion velocity $(k_0)_G=K-C=2.2\pm1.8$~km s$^{-1}$ kpc$^{-1}$ do not differ significantly from zero. From the solution shown in the lower part of the table we find the angular residual rotation velocity
$\Omega_G=6.9\pm0.2$~km s$^{-1}$ kpc$^{-1}$; its sign indicates the rotation in the direction opposite
to that of the galaxy. The angular expansion velocity $(k_0)_G=K-C$ here does not differ significantly from
zero.

Note that we based our choice of the turn angle $l_{xy}$ and $l_{xy+K}$ on which effect was dominant: expansion
or rotation. For example, in the upper part of the table for Step IV we give the results obtained when
turning the coordinate system by angle $l_{xy}=-37^\circ.$ This value is of interest, as it points in the direction $143^\circ$--$323^\circ.$ The center of the ``doughnut'' mentioned above, i.e., the geometric Gould belt center, is located at approximately $\sim140^\circ$ (Fig. 3b). In the lower part of Table 2, Step III does not have this selection problem, since both angles are roughly the same.

As can be seen from equations (11)--(13), the rotation parameters are best determined from equation
(12), which has no $\sin b$ for the unknown $A,$ $B$ and $C$. And vice versa, the expansion/contraction
parameters $K$ and $C$ are best determined from the radial velocities (equation (11)). Therefore, the expansion
parameters in the upper part of Table 2 are more reliable, whereas the rotation parameters are
best taken from the lower part of the table.

Parameters $B$ and $K$ of rotation and expansion, derived in Step II, are in agreement with the
results of other authors. Both in the upper and lower parts of Table 2, we have a positive value of
$K\sim6\pm1$~km s$^{-1}$ kpc$^{-1}$ and a negative $B\sim-3\pm1$~km s$^{-1}$ kpc$^{-1}$. For instance, as Torra et al. (2000) found from an analysis of OB stars younger than 30 million years and located not further than 0.6 kpc from the Sun, $K=7.1\pm1.4$~km s$^{-1}$ kpc$^{-1}$. These authors have not specifically determined $A_G$ and $B_G$, but obtained for the galactic rotation
$A=-8.5\pm2.7$~km s$^{-1}$ kpc$^{-1}$, $B=-24.5\pm2.7$~km s$^{-1}$ kpc$^{-1}$,
$C=10.5\pm2.7$~km s$^{-1}$ kpc$^{-1}$ and $K=7.4\pm2.7$~km s$^{-1}$ kpc$^{-1}$.
Here the strong difference of $B$ from $-15$~km s$^{-1}$ kpc$^{-1}$, typical for galactic rotation, is interpreted as a presence of a noticeable negative intrinsic rotation of the Gould belt. And, of course, no one has previously separated the Scorpius–Centaurus association stars from the Gould belt stars.

 \section{CONCLUSIONS}
We studied the spatial and kinematic properties of young T Tauri type stars from the work of Zari et al.
(2018). The proper motions and parallaxes of such stars were selected from the Gaia DR2 catalog. These
authors also collected the radial velocities for a small percentage of these stars from the literature.

The paper of Zari et al. (2018) describes four ums stellar samples (this is a sample of stars in the upper
part of the Main Sequence): pms1, pms2 and pms3, selected according to their transverse velocities. In
this work, our main attention was directed at analyzing two samples, namely pms1 and pms3. The pms1 sample contains 43 719 T Tauri type stars, but compared to samples pms2 and pms3, it includes the most background objects. The pms3 sample contains 23 686 T Tauri type stars, which are probable Gould
belt members, according to the principle of their selection.

We used the pms1 sample stars to find the following exponential distribution parameters: average
$(z_G)_\odot=-25\pm5$~pc, and scale height $h_G=56\pm6$~pc. We propose using these parameters to eliminate
the background stars located at large heights with respect to the Gould belt symmetry plane. To this end,
we suggest switching to a primed coordinate system $l',b'$ $(x',y',z')$, connected with the Gould belt symmetry
plane. This approach allows one to eliminate all objects that are tilted relative to the Gould belt
symmetry plane, with the mixed stellar composition remaining only in the nodes.

Using about 1800 stars from the pms3 sample with measured radial velocities, proper motions and
parallaxes we found the angular galactic rotation velocity
$\Omega_0=B-A=-30.6\pm1.2$~km s$^{-1}$ kpc$^{-1}$ which is rather close to the known estimates of this
quantity.

We then formed the residual velocities for the pms3 sample stars, free of the Sun’s peculiar velocity
and differential galactic rotation (Step II). This step shows that the stars move with velocity
$5.72\pm0.23$~km s$^{-1}$  in the direction of $l=185\pm2^\circ$ and $b=-12\pm3^\circ$ relative to the local standard of rest. The velocity vector thus lies practically in the Gould belt plane. Among the other parameters, the most noticeable is $K_G=6\pm1$~km s$^{-1}$ kpc$^{-1}$ (the so-called K-effect, which describes the expansion/contraction of a stellar system). In the next stage (Step III and Step IV) we show that the kinematic $K$-effect practically totally disappears when Scorpius-Centaurus association stars are eliminated
from the sample. The angular velocity of intrinsic residual rotation here is small.

Simultaneously, we analyzed the entire pms3 sample, containing more than 23 000 stars with mostly known proper motions and parallaxes; about 1800 of these stars have measured radial velocities. Note that the $K$-effect is present, however, unlike with the previous sample, here we derive in Step I the velocity $U_\odot=10.74\pm0.04$~km s$^{-1}$, which differs from the one obtained earlier by about 5 km s$^{-1}$. This
influences the determination of the residual motion of the sample with regard to the local standard of rest.
The Oort constant $A$ also differs significantly. The last step shows that the residual intrinsic rotation
angular velocity of the pms3 sample amounts to $\Omega_G=6.9\pm0.2$~km s$^{-1}$ kpc$^{-1}$, and this rotation is in the opposite direction to galactic rotation.

 \subsubsection*{ACKNOWLEDGMENTS}
The authors are grateful to the reviewer for useful remarks that helped improve the paper.

 \subsubsection*{FUNDING}
This work has been partially supported by the Program KP19--270 of the Presidium of the Russian
Academy of Sciences ``Questions of the origin and evolution of the Universe with the application
of methods of ground-based observations and space research''.

 \subsubsection*{CONFLICT OF INTEREST}
The authors declare no conflict of interest.

 \bigskip
 \bigskip\medskip{\bf REFERENCES}
{\small

1. The HIPPARCOS and Tycho Catalogues, ESA SP--1200 (1997).
 https://vizier.u-strasbg.fr/viz-bin/VizieR?-source=I/239.

2. V. V. Bobylev, Astronomy Letters 32 (12), 816 (2006).

3. V. V. Bobylev, Astronomy Letters 42 (8), 544 (2016).

4. V. V. Bobylev, Astronomy Letters 64 (2020) [in press].

5. V. V. Bobylev and A. T. Bajkova, Astronomy Letters 42 (1), 1 (2016a).

6. V. V. Bobylev and A. T. Bajkova, Astronomy Letters 42 (3), 182 (2016b).

7. V. V. Bobylev and A. T. Bajkova, Astronomy Letters 45 (4), 208 (2019).

8. V. V. Bobylev and A. T. Bajkova, Astronomy Reports 64 (4), 326 (2020a).

9. V. V. Bobylev and A. T. Bajkova, Astronomy Reports 64 (4), 326 (2020b).

10. C. Bonatto, L. O. Kerber, E. Bica, and B. X. Santiago, Astron. and Astrophys. 446 (1), 121 (2006).

11. Gaia Collab., A. G. A. Brown, A. Vallenari, et al., Astron. and Astrophys. 616, 1 (2018).

12. T. Camarillo, V. Mathur, T. Mitchell, and B. Ratra, PASP 130 (984), 024101 (2018).

13. T. Cantat-Gaudin, C. Jordi, A. Vallenari, et al., Astron. and Astrophys. 618, 93 (2018).

14. T. M. Dame, D. Hartmann, and P. Thaddeus, Astrophys. J. 547 (2), 792 (2001).

15. T. M. Dame, H. Ungerechts, R. S. Cohen, et al., Astrophys. J. 322, 706 (1987).

16. F. Damiani, L. Prisinzano, I. Pillitteri, et al., Astron. and Astrophys. 623, 112 (2019).

17. P. T. de Zeeuw, R. Hoogerwerf, J. H. J. de Bruijne, et al., Astron. J. 117 (1), 354 (1999).

18. Yu. N. Efremov, Centers of star formation in galaxies (Nauka, Moscow, 1989) [in Russian].

19. F. Elias, J. Cabrera-Ca\~no, and E. J. Alfaro, Astron. J. 131 (5), 2700 (2006).

20. J. A. Frogel and R. Stothers, Astron. J. 82, 890 (1977).

21. G. A. Gontcharov, Astronomy Letters 45 (9), 605 (2019).

22. Y. C. Joshi, MNRAS 378 (2), 768 (2007).

23. M. Kounkel and K. Covey, Astron. J. 158 (3), 122 (2019).

24. P. O. Lindblad, Bull. Astron. Inst. Netherlands 19, 34 (1967).

25. P. O. Lindblad, Astron. and Astrophys. 363, 154 (2000).

26. L. Lindegren, J. Hern\'andez, A. Bombrun, et al., Astron. and Astrophys. 616, 2 (2018).

27. G. Marton, P. \'Abrah\'am, E. Szegedi-Elek, et al., MNRAS 487, 2, 2522
(2019).

28. K. F. Ogorodnikov, Dynamics of stellar systems (Pergamon, Oxford, 1965).

29. C. A. Olano, Astron. J. 121 (1), 295 (2001).

30. G. N. Ortiz-Le\'on, L. Loinard, S. A. Dzib, et al., Astrophys. J.869 (2), L33 (2018).

31. P. P. Parenago, Kurs zvezdnoi astronomii (Gosizdat, Moscow, 1954) [in Russian].

32. C. A. Perrot and I. A. Grenier, Astron. and Astrophys. 404, 519 (2003).

33. M. A. C. Perryman, L. Lindegren, J. Kovalevsky, et al., Astron. and Astrophys. 500, 501 (1997).

34. A. E. Piskunov, N. V. Kharchenko, S. R\"oser, et al., Astron. and Astrophys. 445 (2), 545 (2006).

35. W. P\"oppel, Fundamental of Cosmic Physics 18, 1 (1997).

36. A. S. Rastorguev, N. D. Utkin, M. V. Zabolotskikh, et al., Astrophysical Bulletin 72 (2), 122 (2017).

37. B. C. Reed, Astron. J. 120 (1), 314 (2000).

38. M. J. Sartori, J. R. D. L\'epine, and W. S. Dias, Astron. and Astrophys. 404, 913 (2003).

39. E. F. Schlafly, G. Green, D. P. Finkbeiner, et al., Astrophys. J. 786 (1), 29 (2014).

40. R. Sch\"onrich, J. Binney, and W. Dehnen, MNRAS 403 (4), 1829 (2010).

41. C. Soubiran, T. Cantat-Gaudin, M. Romero-G\'omez, et al., Astron. and Astrophys. 619, 155 (2018).

42. R. Stothers and J. A. Frogel, Astron. J. 79, 456 (1974).

43. J. Torra, D. Fern\'andez, and F. Figueras, Astron. and Astrophys. 359, 82 (2000).

44. T. N. G. Westin, Astron. and Astrophys. Suppl. 60, 99 (1985).

45. E. Zari, H. Hashemi, A. G. A. Brown, et al., Astron. and Astrophys. 620, 172 (2018).
 }

  \end{document}